\documentclass[pra,aps,amsmath,amssymb,showpacs,showkeys]{revtex4}
\usepackage{graphicx}
\usepackage{epsfig}
\usepackage{epstopdf}
\usepackage{color}

\newcommand{\bla}{\color{black}}

\begin{document}

\title{Two-step orthogonal-state-based protocol of quantum
  secure  direct communication  with the  help of  order-rearrangement
  technique}        {\color{black}        \author{Preeti        Yadav}
  \email{pri8.phy@gmail.com}  \affiliation{Dept. of Physics,
    Indian Institute of Technology Kanpur,  Kanpur- 208016, India.}   \author{R.  Srikanth}
  \email{srik@poornaprajna.org} \affiliation{Poornaprajna Institute of
    Scientific  Research,  Sadashivnagar,  Bengaluru-  560080,  India}
  \affiliation{Raman  Research  Institute,  Sadashivnagar,  Bengaluru-
    560060,          India.}           \author{Anirban         Pathak}
  \email{anirban.pathak@jiit.ac.in}  \affiliation{Jaypee  Institute of
    Information Technology, A-10,  Sector-62, Noida, UP-201307, India}
  \affiliation{RCPTM, Joint Laboratory of Optics of Palacky University
    and \\  Institute of  Physics of Academy  of Science of  the Czech
    Republic, Faculty  of Science, Palacky  University, 17.  listopadu
    12, 77146 Olomouc, Czech Republic.}

\begin{abstract}
The  Goldenberg-Vaidman  (GV) protocol  for  quantum key  distribution
(QKD)  uses orthogonal encoding  states of  a particle.   Its security
arises  because  operations  accessible  to Eve  are  insufficient  to
distinguish  the two  states encoding  the secret  bit.  We  propose a
two-particle  cryptographic protocol   for  quantum secure
direct  communication, wherein  orthogonal  states encode
the secret,  and security arises  from restricting Eve  from accessing
any  two-particle   operations.   However,  there   is  a  non-trivial
difference  between the  two  cases.  While  the  encoding states  are
perfectly indistinguishable in  GV, they are partially distinguishable
in the bi-partite  case, leading to a qualitatively  different kind of
information-vs-disturbance trade-off  and also options for  Eve in the
two cases.
\end{abstract}


\maketitle

\section{\label{sec:Introduction}Introduction}

Recent    advances   in   device-independent    quantum   cryptography
\cite{mpa11} have brought to  the fore the relevance of multi-particle
systems  in quantum cryptography,  following a  line of  thought first
initiated  by  Ekert  \cite{E91}.   In  response to  this  work,  Ref.
\cite{bbm92} proposed a quantum key distribution (QKD) scheme based on
Einstein-Podolsky-Rosen  (EPR) correlations,  which was  equivalent to
the  original Bennett-Brassard  1984 (BB84)  \cite{bb84}  protocol for
QKD,  but uses  separable particles  instead of  entangled  ones.  The
argument  of  Ref.   \cite{bbm92}  would  suggest  that  the  security
features of  EPR were  reflected in BB84.   A similar  relation exists
arguably between  the Ping-pong \cite{ping-pong} on the  one hand, and
DL04-QSDC  \cite{DL04} or  Lucamarini-Mancini 2005  (LM05) \cite{lm05}
protocols,  on  the  other,  in  the  sense that  the  former  may  be
considered as the entangled version of the latter.

All the separable-state protocols discussed above, BB84, DL04-QSDC and
LM05, employ non-orthogonal states, whose perfect indistinguishability
lies   at   the   heart   of   their   security.    Further,   perfect
indistinguishability of  non-orthogonal states also  provides security
to many  other protocols of QKD,  such as B92  \cite{b92} and DL04-QKD
\cite{Deng-Long04}  protocols.  By  contrast,  Goldenberg and  Vaidman
\cite{GV95}  proposed  a  protocol  (GV),  demonstrating  that  secure
cryptography can be accomplished even with orthogonal states.  The key
point was  that they  were superpositions of  geographically separated
wave  packets.  Secrecy  in this  case arises  because of  the  set of
operations Eve can apply are restricted by quantum mechanics.

Most   of   the  early   protocols   \cite{bb84,b92,E91}  of   quantum
cryptography  were  limited   to  QKD.   Specifically,  these  quantum
cryptographic  protocols are designed  to generate  an unconditionally
secure    random    key  by  quantum  means  and  subsequently
classical  cryptographic procedures  are  used to  encode the  message
using the key generated  by these protocols.  Interestingly, later
protocols  for secure  communiation \cite{banerjee2012,long2007review}
were proposed  that allow to either  generate a \textit{deterministic}
key or to circumvent the  prior generation of key.  These protocols of
secure direct quantum communication  can broadly be divided into three
sub-classes:  (i) Deterministic  QKD protocols  \cite{GV95, ping-pong,
  lm05}; (ii) protocols for deterministic secure quantum communication
(DSQC) \cite{banerjee2012,  xiu2009DSQC, long2007review}; and finally,
(iii)  protocols  for   quantum  secure  direct  communication  (QSDC)
\cite{DL04}.

In deterministic  QKD and DSQC,  there is some information  leakage of
classical   data  prior   to  detection   of  eavesdropping   by  Eve.
Deterministic QKD  solves this problem  by transmitting a  random key,
rather  than the  secret message.   In  DSQC, the  receiver (Bob)  can
decode the  message only  after receipt of  an encoding key,  which is
some additional classical information (at least one bit for each qubit
transmitted  by the  sender (Alice)).   Thus in  the event  of leakage
detection, the encoding key is  not published, in order to protect the
message.

In contrast to DSQC, when  no such additional classical information is
required, a direct secure quantum  communication of the message can be
achieved, which  happens in  a QSDC protocol.   Protocols of  DSQC and
QSDC  are interesting  for various  reasons.  Firstly,  a conventional
QKD-based quantum communication protocol uses a classical intermediate
step to  transmit the message,  but no such classical  intermediary is
required in  DSQC and QSDC\bla. Further,  a QSDC or  DSQC protocol can
always be turned  into a protocol of QKD as the  sender who is capable
of communicating a meaningful message can also choose to communicate a
random string of bits to convert  the protocol into a protocol of QKD.
However, the converse is not true (i.e., a QKD protocol cannot be used
as a protocol of QSDC or DSQC).

In  this   work,  we   consider  an  orthogonal-state   based  quantum
cryptography  protocol   that  uses  two-particle   entanglement.   By
transmitting the two particles  separately, we obtain security because
the set of states distinguishable via the accessible operations to Eve
fail  to distinguish  the encoding  states.   As in  GV, our  protocol
requires delayed measurement  on the first particle in  order to work.
(Therefore, from a practical perspective, quantum memory, an expensive
resource, is  required).  By contrast,  for non-orthogonal-state based
protocols,  delayed  measurement  is  replaced by  random  measurement
choice,  but can  be  used to  improve  efficiency \cite{DLW+04}.   An
important difference  with the single-particle  case is the  degree of
distinguishability, thus making the  proof of security quite different
in  the two-particle  case.    Further, our  use  of block
transmission and  an order-rearrangement technique  makes the protocol
suitable for QSDC, while GV is a protocol for QKD. 

In  GV, both  the encoding  states  as well  as error-checking  states
involve only  orthogonal states, while  in BB84, both types  of states
are  non-orthogonal.  More  generally, one  may  consider cryptography
protocols that  involve orthogonal-state encoding  but allow conjugate
coding   for  error-checking   \cite{LL02,   BGL+04,  WDL+05,   WDL05,
  LDZ+08}. Using  the strategy  adopted for eavesdropping  checking in
the protocol  proposed in the present  paper it is  possible to modify
these  protocols  \cite{LL02,  BGL+04,  WDL+05,  WDL05,  LDZ+08}  into
equivalent completely orthogonal-state-based protocols.

\section{\label{sec:GV}The GV protocol and its security}

We briefly review GV. Alice and Bob are located at two ends of a large
Mach-Zehnder interferometer.  Let $|U(t)\rangle$ and $|L(t)\rangle$ be
two localized wave  packets of Alice's particle $S$,  travelling by the
upper and  lower arm  of the interferometer,  respectively.  Classical
bit $j (= 0,1)$ is encoded as:
\begin{equation}
|\Psi_{j}\rangle  =     \frac{1}{\sqrt{2}}\left(|U(t_s)\rangle       +
(-1)^j|L(t_s)\rangle\right),
\label{eq:psij}
\end{equation}
where it is  assumed that there is no overlap  between the supports of
$\{|U(t)\rangle\}$  and $\{|L(t)\rangle\}$.   Alice  sends Bob  either
$|\Psi_{0}\rangle$  or $|\Psi_{1}\rangle$  by delaying  packet  $L$ by
time  $\Delta$   to  ensure  that  $|U\rangle$   and  $|L\rangle$  are
\textit{not} present in the channel at the same time.  In his station,
Bob receives $|U(t_s+\tau)\rangle$, where $\tau$ is the travel time of
the pulse  from Alice's to Bob's  station. Bob puts the  pulse on hold
for time  $\Delta$ (where $\tau  < \Delta$), before combining  it with
$|L(t_s+\Delta+\tau)\rangle$,  to  recreate  the  superposition  state
$|\Psi_j^\prime\rangle$,  which  is  the same  as  $|\Psi_{j}\rangle$,
apart from  an inconsequential global  phase. Bob then  decodes bit
$j$ deterministically from his interferometric output.

Alice and Bob perform the following two tests to detect Eve's possible
malicious  eavesdropping: (1)  They compare  the sending  time $t_{s}$
with the  receiving time  $t_{r}$ for each  wave packet. We  must have
$t_{r}=t_{s}+\tau+\Delta.$  This  ensures that  Eve  cannot delay  the
upper  packet until  also  having lower  packet simultaneously,  which
would allow her  to decode the states.  Even so,  she may replace both
wave packets with a corresponding dummy.  To avoid such an attack, the
timing of  transmission of particles  is kept random.   To faciliatate
this, Alice and Bob discretize  their sending times into a sequence of
time bins.   (2) Alice selects  a fraction of particles  and announces
their  time coordinates.   Bob announces  his measurement  outcomes on
them.  Alice  ensures that his  received bits are consistent  with her
transmitted bits.

The  security  of  GV  can  be  understood in  terms  of  an  extended
no-cloning  theorem  applicable   to  orthogonal  states,  when  Eve's
operations are restricted  by the fact that she  can physically access
only one  of the pieces  $|U(t)\rangle$ and $|L(t)\rangle$ at  a given
time \cite{M98}.  We present a slightly different version, amenable to
subsequent generalization.

The simplest operation accessible is a projective measurement onto the
basis  $\{|U(t)\rangle,|L(t)\rangle\}$,   where  we  may   ignore  the
time-dependence for convenience.  If Eve measures projectively in this
basis, she merely disrupts the  coherence between the wave packets and
is  detected, but  obtains no  information about  the secret  bit $j$.
More  generally, Eve  can introduce  a probe  $P$ that  interacts with
Alice's particle according to:
\begin{equation}
\mathcal{U} \equiv |U\rangle\langle  U| \otimes C_U + |L\rangle\langle
L|\otimes C_L,
\label{eq:mez}
\end{equation}
where  $C_U$ and  $C_L$ are  unitaries  acting on  the ancilla  alone.
Because the two  packets are never together on  the channel, causality
demands  that  Eve's attack  cannot  unitarily  mix  the $U$  and  $L$
pieces. An implication is that no  attack by Eve, which is confined to
the form (\ref{eq:mez}),  can extract secret bit $j$,  because this is
stored  as the  phase information  between the  two wave  packets, and
cannot  be accessed  \textit{even probabilistically}.   We  prove this
below.

Let $|R\rangle$ be the initial  `ready' state of the probe.  Acting on
the particle-probe  system, Eq.  (\ref{eq:mez})  transforms an initial
state $|\Psi_j\rangle \otimes |R\rangle$, after they are recombined by
Bob, to the state
\begin{equation}
\rho_{SP}   =   \frac{1}{4}\left(
\begin{array}{cc}
|u\rangle\langle    u|  &   (-1)^j |d\rangle\langle u| \\
 ~ & ~ \\
(-1)^j|u\rangle\langle     d|  & 
|d\rangle\langle  d| 
\end{array}\right),
\end{equation}
where $|u\rangle \equiv  C_0|R\rangle$ and $|d\rangle = C_1|R\rangle$.
The probe is now left in the state:
\begin{equation}
\rho_P^\prime           =          \textrm{Tr}_S(\rho_{SP})          =
\frac{1}{2}(|d\rangle\langle d| + |u\rangle\langle u|),
\label{eq:detector}
\end{equation}
which,  as  with  the  case  of  projective  measurements,  yields  no
information  to Eve about  the secret  bit $j$.   In other  words, Eve
gains nothing by  attacking in the case of  individual attacks.  It is
not difficult to  see that this is also true  for Eve's collective and
joint  attacks.  

\begin{figure}
\centering          
\includegraphics[width=9cm]{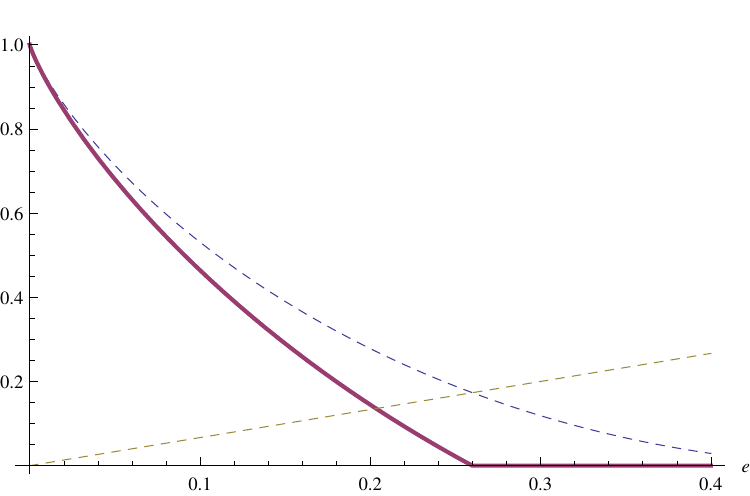}  
\caption{Bob's  information ($I_B$,  falling dashed  curve)  and Eve's
  information  ($I_E$, rising  line), respectively,  as a  function of
  eavesdropping parameters observed error $e$. For $e \geq e_{\rm max}
  \equiv  0.26$, $I_E  > I_B$.  The  falling line  corresponds to  the
  positive key rate (\ref{eq:key}).}
\label{fig:info_dist_GV} %
\end{figure}

Assuming ideal single-photon sources and detectors with Alice and Bob,
the  only  way  for  Eve  to  attack the  GV  protocol,  is  that  she
substitutes  dummies   by  blocking   fraction  $f$  of   the  genuine
particles. Suppose that  Alice and Bob agree to  discretize the random
sending time. In each sequential block of $\gamma$ (an integer) number
of  time steps, one  particle is  transmitted by  Alice in  a randomly
chosen  time cell  within the  block.  Eve's  strategy would  be to fully
blockade a  fraction $f$ of  randomly chosen blocks,  and transmit a
dummy  prepared by  her in  a  randomly chosen  time  within  the block.   The
probability that  she gets a  match with Alice's transmission  cell is
$1/\gamma$.

To calculate the error rate Eve generates, we note that
even when Eve  gets the timing right,
she will  be wrong  half the  time about the  encoded state.  Thus Eve
generates error rate
\begin{equation}
e  = f \times 
\left[\frac{1}{2}\frac{1}{\gamma} + \left(1 - \frac{1}{\gamma}\right)\right]
= f \times \left(1-\frac{1}{2\gamma}\right), 
\label{eq:wr}
\end{equation}
where $\gamma$  is a publicly known number.  Bob's average information
on the  sifted bits  is given $I_B  \equiv I(A:B)  = 1 -  h(e)$, where
$h(\cdot)$ is  the binary Shannon  information.  On the  dummies whose
timing is right, Eve has full information, i.e.,
\begin{equation}
I(A:E)  =  I(B:E)  \equiv  I_E= \frac{f}{\gamma}
= \frac{2e}{2\gamma-1},
\label{eq:Evef}
\end{equation}  
where the last equation follows from Eq. (\ref{eq:wr}) and
the  $I(A:E)$ and $I(B:E)$ denote Eve's  mutual information on
Alice and  Bob.  She knows  when she got  it right when Alice  and Bob
perform the equivalent of basis reconciliation for the sending times.

The  corresponding data is  plotted in  Figure \ref{fig:info_dist_GV}.
The  requirement  for  positive  secret  key  rate  is  determined  by
\cite{CK78}
\begin{equation}
K \equiv I_B - \textrm{min}(I(A:E),I(A:B)) = I_B-I_E,
\label{eq:key}
\end{equation}
from which  the maximum  tolerable error is  found to be  $e_{\rm max}
\approx 0.26$.  

\section{Towards a two-particle orthogonal-state based 
protocol \label{sec:2multi}}

In  seeking  a  protocol  that   extends  GV  to  a  two-particle  (or
multi-particle)   scenario,   we  are   naturally   led  to   consider
cryptographic adaption  of quantum  dense coding to  cryptography (cf.
the protocol of Ref. \cite{Deng}  for dense coding based secure direct
communication.)  On  the analogy  of GV, one  might expect  that Alice
should  transmit the  two  entangled particles  one  after another  at
random timings and  such that both are not found  on the open channel.
Surprisingly,  this  can be  completely  insecure  against Eve,  whose
strategy  would be  as follows.   When the  first particle  comes, she
holds it, and transmits her own half of a Bell state towards Bob.  She
can in  principle find  out the position  of the randomly  sent second
particle, measure  it jointly  with Alice's first  particle, determine
their joint  state, and then  transmit a dummy  particle appropriately
entangled with  her first  dummy particle. Here  we have  assumed that
Eve's measurements take negligible time. 

To avoid this attack,  such bi-partite cryptographic protocols may add
multi-partite  non-orthogonal states either  to the  coding or  in the
checking  step (as  in BB84).   However,  if we  remain restricted  to
orthogonal states, then the order  of particles needs to be scrambled,
via the permutation of particle (PoP) action, an idea first introduced
by  Deng and  Long in  2003 in  a pioneering  work \cite{DL03}  on the
``controlled order rearrangement encryption'' (CORE) QKD protocol.
In  the present  work, a  two-particle QSDC  protocol inspired  by GV,
which is  referred to as  2GV, is presented  along these lines  in the
next section.

Now  suppose   Eve  does   \textit{not}  launch  the   dummy  particle
attack.  Assuming   ideal  sources   and  detectors,  GV   is  secure.
Interestingly, a bipartite generalization  of GV (without PoP) is not.
The reason is interesting  and highlights a difference between single-
and  bi-partite nonlocality:  while  Eve gets  no  information on  the
encoded  bits  when  the  two  packets  are  de-synchronized,  in  the
bipartite  case,  partial information  can  be  obtained, as  detailed
below.

Alice  and Bob employ  a key  distribution protocol  where the  key is
shared via  a dense coding strategy,  and must test for  Eve after the
transmissions are completed.  Alice and Bob need to model Eve's attack
strategy, and  estimate whether Eve's information  on their secret
bits, as a function of observed noise, is too high to be eliminated by
subsequent  classical post-processing.  If  it is,  only then  do they
abort the protocol run.  We  furnish a security proof of the protocol,
assuming  individual  attacks  by  Eve  on  each  of  the  two  coding
particles.   From   this  we  extract   an  information-vs-disturbance
trade-off, and hence determine the largest tolerable error rate.

As a  specific example of  the attack employed  by Eve, we  consider a
model given in Ref.  \cite{zbi}, which is based on one proposed by Niu
and Griffiths \cite{NiuGri}.  Probes  $E_{0}$ an $E_{1}$ interact with
each     transmitted     qubit,     being     subjected     to     the
interaction:  \begin{eqnarray}   |0\rangle|E\rangle  &  \rightarrow  &
  \sqrt{\frac{1+\cos\theta}{2}}|0\rangle|\epsilon_{0}\rangle+
  \sqrt{\frac{1-\cos\theta}{2}}         |1\rangle|E_{0}\rangle\nonumber
  \\         |1\rangle|E\rangle         &        \rightarrow         &
  \sqrt{\frac{1+\cos\theta}{2}}|1\rangle|\epsilon_{1}\rangle+
  \sqrt{\frac{1-\cos\theta}{2}}|0\rangle|E_{1}\rangle,
\label{eq:niugri}\end{eqnarray}
where,  furthermore   $\langle\epsilon_0|\epsilon_1\rangle  =  \langle
E_0|E_1\rangle  = \cos\theta$  by  virtue of  symmetry  in the  attack
strategy.   For  simplicity, the  same  attack  parameter $\theta$  is
assumed to characterize the attack on both particles.  This results in
the initial state $\rho_{AB}$, which  is a Bell state in 2GV, evolving
into    a    joint    state    of   the    particles    and    probes,
$\rho_{ABE_1E_2}^{\prime\prime}$.

After some straightforward calculation,  the above attack can be shown
to produce the reduced density operator
\begin{equation}
\rho_{AB}^{\prime\prime}             =            \textrm{Tr}_{E_1E_2}
\left(\rho_{ABE_1E_2}^{\prime\prime}\right)                           =
\left(  \begin{array}{cccc} \frac{1}{2}(1 +  \cos^2\theta) &  0 &  0 &
  \frac{1}{2}(1     +     \cos^2\theta)\cos^2\theta     \\     0     &
  \frac{1}{2}\sin^2\theta  &  \frac{1}{2}\sin^2\theta\cos^2\theta &  0
  \\ 0 & \frac{1}{2}\sin^2\theta\cos^2\theta & \frac{1}{2}\sin^2\theta
  &  0  \\  \frac{1}{2}(1  +  \cos^2\theta)\cos^2\theta  &  0  &  0  &
  \frac{1}{2}(1 + \cos^2\theta)
\end{array}\right).
\end{equation}
The quality of state received by Bob can be quantified by the fidelity
$\langle\Phi^+|\rho_{AB}^{\prime\prime}|\Phi^+\rangle     =    (1    +
\cos^2\theta)^2$      (where      we      assume     $\rho_{AB}      =
|\Phi^+\rangle\langle\Phi^+|$).  It  follows that in  order to produce
no  errors,  Eve must  ensure  that  $\theta=0$,  which by  virtue  of
Eq. (\ref{eq:niugri}), implies that  no entanglement is generated, and
in fact $|\epsilon_0\rangle  = |\epsilon_1\rangle$, implying a trivial
interaction of  the probe with Alice's  qubit. Thus, if  no errors are
generated,  then Eve  gains  no information.  More generally,  suppose
finite errors are observed.

\begin{figure}
\begin{tabular}{cc}
\centering          \includegraphics[width=7cm]{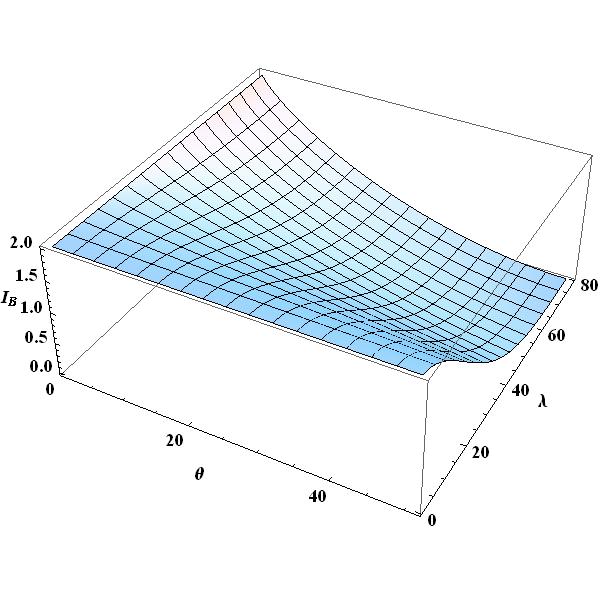}         &
\includegraphics[width=7cm]{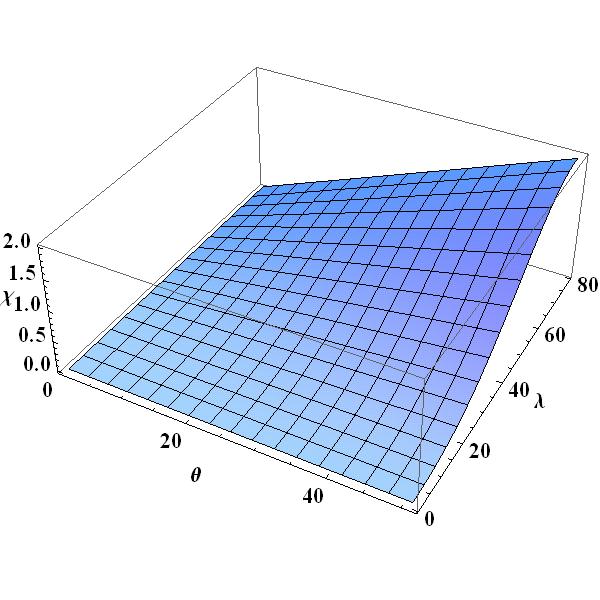} \tabularnewline
\end{tabular}
\caption{(A) Bob's information as  a function of
  eavesdropping  parameters   $\theta$  (overlap  angle,   defined  by
  Eq.  (\ref{eq:niugri})) and  $\lambda$ (fraction  of  particle pairs
  attacked); (B) Eve's information as a function of the same parameters. }
\label{fig:BobEvesInfo} %
\end{figure}

The error rate observed by Alice and Bob is given by: 
\begin{equation}
  e=1-\langle\Phi|\rho^{\prime}_{AB}|\Phi\rangle
\label{eq:error}
\end{equation}
where       $|\Phi\in\{\Phi^{\pm}\rangle,|\Psi^{\pm}\rangle\}$      and
\begin{equation}
\rho^{\prime}_{AB}(\theta,\lambda) = (1-\lambda)|\Phi\rangle\langle\Phi|
 + \lambda \rho^{\prime\prime}_{AB}
\label{eq:rhopab}
\end{equation}
is the corresponding two-particle  state obtained assuming Eve attacks
fraction $\lambda$  of the incoming particle  pairs with eavesdropping
parameter  $\theta$  as  defined  in  Eq.   (\ref{eq:niugri}).   Bob's
information $I_{B}$ is quantified  as the Alice-Bob mutual information
$I_B \equiv I(A:B)$ when Bob measures the incoming states in the Bell
basis. As a function of $\theta, \lambda$, it is:
\begin{equation}
I_B(\theta,\lambda) = H(A) - H(A^\prime(\theta,\lambda)|B = \Phi),
\label{eq:IBE}
\end{equation}
where    $H(A)$    is    Alice's    preparation    entropy    and    $
H(A^\prime(\theta,\lambda)|B  = \Phi)$ is  the conditional  entropy of
$\rho^\prime_{AB}(\theta,\lambda)$  when  Bob  measures  in  the  Bell
basis. The quantity is  presented in Fig.  \ref{fig:BobEvesInfo}A as a
function of Eve's attack parameters.

Eve's information $I_E \equiv I(A:E) = I(A:B)$ is upper-bounded by the
Holevo bound $\chi$ of the reduced density operator of the two probes:
\begin{equation}
\chi = S\left(\sum_j p_j \rho^{\prime(j)}_{E_1E_2}\right)
- \sum_j S\left(\rho^{\prime(j)}_{E_1E_2}\right) \ge I_E(\theta,\lambda),
\label{eq:holevo}
\end{equation}
where $\rho^{\prime(j)}_{E_1E_2}$  ($j=0,1,2,3$) is the  noisy version
of  the  density  operator  corresponding  to  the  four  Bell  states
$|\Phi^\pm\rangle,  |\Psi^\pm\rangle$,  respectively,  being  sent  by
Alice.   This   bound  on  Eve's  information  is   depicted  in  Fig.
\ref{fig:BobEvesInfo}B.   Using  the  following  notation:  $c  \equiv
\cos(\theta), s \equiv \sin(\theta),  K \equiv \frac{1}{2}(1 + c)$, $A
\equiv  K^2c^2s^2$, $B \equiv  K^2c^3s, C  \equiv K^2s^3c$,  $D \equiv
\frac{1}{4}s^3c^3$,   $E    \equiv   \frac{1}{4}s^4c^2$,   $F   \equiv
\frac{1}{4}s^3c^2$,   $H    \equiv   K^2(1   +    c^4)$,   $I   \equiv
\frac{1}{4}(1+c^2)s^2c$,   $J  \equiv   \frac{1}{4}s^4c$,   $L  \equiv
\frac{1}{4}s^2(1  +  c^4)$,   $M  \equiv  \frac{1}{4}s^3$,  $N  \equiv
\frac{1}{4}s^5c$, $P \equiv (1 -  K)^2cs$, $Q \equiv (1 - K)^2s^2$, $R
\equiv  2(1  -  K)^2c^2$,  we  find that  if  Alice  transmits  states
$|\Phi^\pm\rangle$, then the corresponding probe states of Eve are:
\begin{equation}
\rho^\pm_{E_1E_2} =
\left(\begin{array}{cccccccccccccccc}
H & B & 0 & 0 & B & A & 0 & 0 & 0 & 0 & \pm I & \pm M & 0 & 0 & \pm F & 0 \\
B & A & 0 & 0 & A & C & 0 & 0 & 0 & 0 & \pm F & 0 & 0 & 0 & \pm J & 0 \\
0 & 0 &  L & D & 0 & 0 & D & E & \pm I & \pm M & 0 & 0 & \pm F & 0 & 0 & 0 \\
0 & 0 & D & E & 0 & 0 & E &  N & \pm F & 0 & 0 & 0 & \pm J & 0 & 0 & 0 \\
B & A & 0 & 0 & A & C & 0 & 0 & 0 & 0 & \pm F & 0 & 0 & 0 & \pm J & 0 \\
A & C & 0 & 0 & C & K^2s^4 & 0 & 0 & 0 & 0 & \pm J & 0 & 0 & 0 & \pm\frac{1}{4}s^5 & 0 \\
0 & 0 & D & E & 0 & 0 & E &  N & \pm F & 0 & 0 & 0 & \pm J & 0 & 0 & 0 \\
0 & 0 &E &  N & 0 & 0 &  N & \frac{1}{4}s^6 & \pm J & 0 & 0 & 0 & \pm\frac{1}{4}s^5 & 0 & 0 & 0 \\
0 & 0 & \pm I & \pm F & 0 & 0 & \pm F & \pm J & \frac{1}{2}s^2c^2 &  Mc & 0 & 0 &  Mc & 0 & 0 & 0 \\
0 & 0 & \pm M & 0 & 0 & 0 & 0 & 0 &  Mc & \frac{1}{4}s^4 & 0 & 0 & 0 & 0 & 0 & 0 \\
\pm I & \pm F & 0 & 0 & \pm F & \pm J & 0 & 0 & 0 & 0 & R & P & 0 & 0 & P & 0 \\
\pm M & 0 & 0 & 0 & 0 & 0 & 0 & 0 & 0 & 0 & P & Q & 0 & 0 & 0 & 0 \\
0 & 0 & \pm F & \pm J &0&0& \pm J & \pm\frac{1}{4}s^5 &  Mc & 0 & 0 & 0 & \frac{1}{4}s^4 & 0 & 0 & 0 \\
0 & 0 & 0 & 0 & 0 & 0 & 0 & 0 & 0 & 0 & 0 & 0 & 0 & 0 & 0 & 0 \\
\pm F & \pm J &0&0&\pm J & \pm\frac{1}{4}s^5 & 0 & 0 & 0 & 0 & P & 0 & 0 & 0 & Q & 0 \\
0 & 0 & 0 & 0 & 0 & 0 & 0 & 0 & 0 & 0 & 0 & 0 & 0 & 0 & 0 & 0
\end{array}\right).
\end{equation}
It  is immediately seen  that $\rho^+_{E_1E_2}$  and $\rho^-_{E_1E_2}$
are not identical,  implying that Eve can gain  some information about
Alice's  transmission by  distinguishing  $\rho^{\pm}_{E_1E_2}$.  This
encoding-dependence of Eve's  probe state in 2GV is  in stark contrast
to  the  general probe  state  (\ref{eq:detector})  obtained when  Eve
attacks GV.  Thus  Eve's attack in 2GV can  obtain partial information
about the  encoding, whereas she  obtains none in  the case of  GV, even
when no dummy states are used.  Therefore, unlike with GV, in the case
of  there is  an  information-vs-disturbance trade-off  even when  Eve
employs no dummy particles. 

The   tolerable   error   rate   is  computed   as:   
\begin{equation}
  e_{0}=\min_{I_{B}-\chi=0}e,
\label{eq:cliff}
\end{equation}  
the smallest  error for  which $\chi$ just  exceeds $I_B$.  It  may be
considered as the problem of  minimizing $e$ subject to the constraint
that  Eve's  information has  zero  excess  over Bob's.   Numerically,
applying      the      criterion      (\ref{eq:cliff})     to      the
information-vs-disturbance      trade-off     data      in     Figures
\ref{fig:BobEvesInfo},   we    found   the   tolerable    error   rate
$e_{0}=26.7\%$, as plotted  in Figure \ref{fig:errorrate}. This rather
large tolerance can be attributed to the limited power of Eve's attack
here.
\begin{figure}
\centering   
\includegraphics[width=7cm]{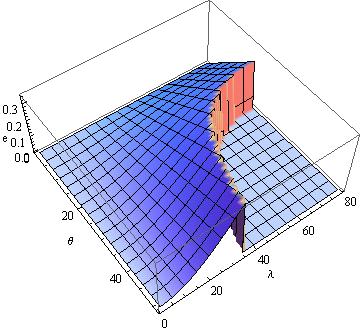}
\caption{Plot of error (\ref{eq:error})  for which a positive key rate
  (\ref{eq:key})   exists   according    to   the   data   of   Figure
  \ref{fig:BobEvesInfo}. This is the  plot of $e = e(\theta,\lambda)$,
  which  is truncated  over the  region  where $I_{AE}  > I_{AB}$.  The
  maximum  tolerable   error  is   the  minimum  across   the  `cliff'
  (cf. Eq. (\ref{eq:cliff})}
\label{fig:errorrate}
\end{figure}

\section{Two-particle orthogonal-state based protocol \label{sec:2GV}}

Instead  of random  transmission, Alice  transmits multiple  halves of
Bell  states,  and scrambles  the  order  of  the second  halves.   In
realistic protocols,  there will be  inevitable noise.  As  is usually
done, any error observed by Alice  and Bob is attributed to a putative
Eve's intervention, though errors can arise also due to channel noise,
too.   The  presented  scheme  enumerated  below  uses  this  idea  of
re-ordering or  permutation of particle  order.  An illustration of the
protocol is given in Figure \ref{fig:2GV_protocol}. 
\begin{figure}
\centering          
\includegraphics[width=9cm]{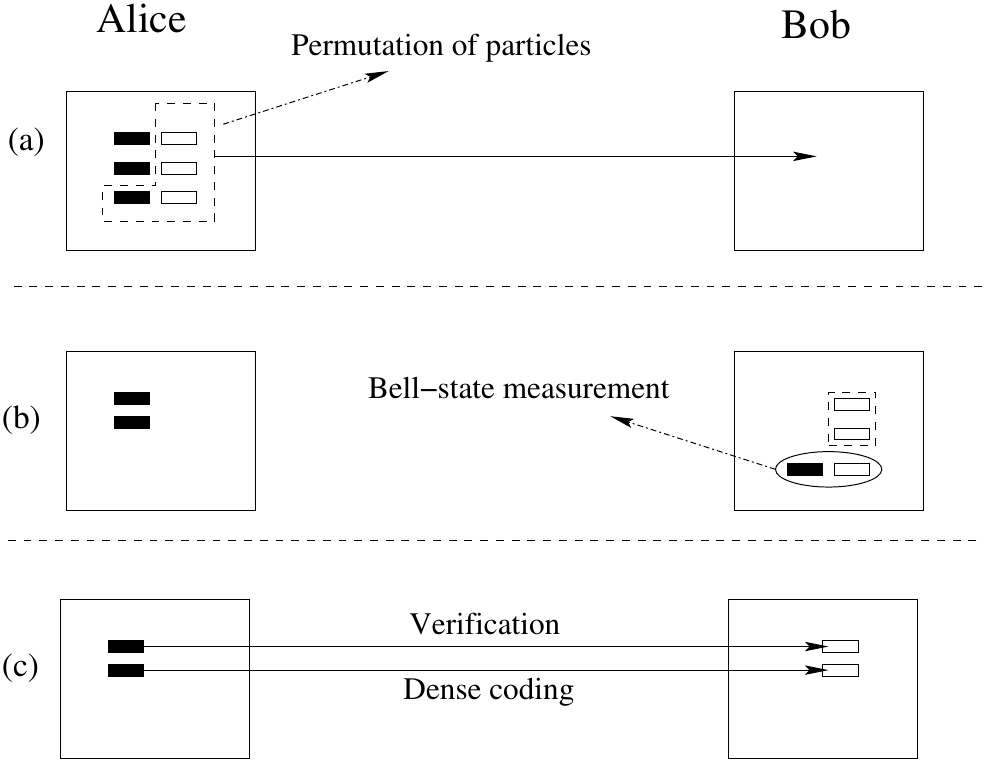}  
\caption{Illustration of  the quantum information  processing steps of
  the new protocol, where we indicate the block transmission of qubits
  (cf. Ref. \cite{LL02})  (a) Step 1 of the  new protocol, where Alice
  transmits qubits of the set  $S_2^{(b)}$ ($2n$) qubits and set $S_1$
  ($2n$  qubits)   to  Bob,  after  permuting   them  using  operation
  $\Pi_{4n}$  (indicated  as  the  dashed  box); (b)  Step  2  of  the
  protocol, where  Bob performs a Bell-state measurement  on the $S_1$
  qubits,  after  reordering  them  based  on  coordinate  information
  received  from  Alice; the  particles  of  $S^{(b)}_2$ still  remain
  permuted (dashed box); (c) Step  3, where Alice sends the $n$ qubits
  used for verification,  and $n$ used for transmitting  a random key,
  along with  the required  coordinate information. After  these three
  steps Alice and Bob collaborate to check eavesdropping in Step 4 and
  finally  in Step  5 Bob  performs  Bell measurements  to obtain  the
  shared key (Steps 4 and 5 are not shown in this figure).}
\label{fig:2GV_protocol} %
\end{figure}

\begin{enumerate}
\item Alice  prepares the state  $|\Psi^{+}\rangle^{\otimes 3n}$ where
  $|\Psi^{+}\rangle = \frac{|00\rangle + |11\rangle} {\sqrt{2}}$. 
  She divides them into two sets: set $S_1$ of $n$ pairs and set $S_2$
  of $2n$ pairs.  Let $S_j^{(a)}$ denote the first  half ($jn$ qubits)
  of set $S_j$,  and $S_j^{(b)}$ denote the second  half of set $S_j$.
  She keeps $S_2^{(a)}$ with  herself ($2n$ qubits).  On the remaining
  $4n$  qubits   of  $S_1  \cup  S^{(b)}_2$,  she   applies  a  random
  permutation operation $\Pi_{4n}$ and  transmits them to Bob; $2n$ of
  the transmitted qubits  are Bell pairs (the members  of $S_1$) while
  the remaining  $2n$ (the members  of $S^{(b)}_2$) are  the entangled
  partners of the particles remaining with Alice. 
\item  After receiving Bob's  authenticated acknowledgment, Alice
  classically announces  the coordinates of the $2n$  members of $S_1$
  among the transmitted particles.   Bob measures them in the Bell
  basis to determine  if they are each in  the state $|\Psi^+\rangle$.
  If  the error  detected by  Bob is  within a  tolerable  limit, they
  continue to the next step.  Otherwise, they discard the protocol and
  restart from Step 1.
\item Alice  randomly chooses  a sequence of  $n$ qubits from  the set
  $S_2^{(a)}$  in  her possession  to  form  the verificiation  string
  $\Sigma_2^{(a|V)}$ for the next  round of communication, and encodes
  her key in the remaining $n$  qubits of $S_2^{(a)}$ to form the code
  string $\Sigma_2^{(a|C)}$.  To encode  a 2-bit message or key, Alice
  applies one of  the 4 Pauli (dense coding) operations  $I, X, iY, Z$
  on her qubit.  After the  encoding operation, Alice sends all qubits
  in her possession (i.e., $S_2^{(a)}$) to Bob.
\item  Alice  discloses the  coordinates  of  the verification  qubits
  ($\Sigma_2^{(a|V)}$)  and their  partner  particles after  receiving
  authenticated acknowledgement of receipt of all the qubits from Bob.
  Bob performs  Bell measurement on the verification  qubits and their
  partner particles and computes the error rate as in Step 2.
\item If  the error  rate is tolerably  low, then Alice  announces the
  coordinates of  the partner  partcles of $\Sigma_2^{(a|C)}$  and Bob
  uses that  information to  decode the encoded  message or key  via a
  Bell-state measurement  on the  remaining Bell pairs,  and classical
  post-processing.
\end{enumerate}

2GV may be  considered as the bi-partite generalization  of GV because
the encoding is via orthogonal states, and security arises because the
encoding states  cannot be distinguished by  the restricted operations
available  to Eve.  However,  there are  three important  differences.
First is,  as noted above, that randomizing  the transmission schedule
of  Alice's   particle  does  not  help.   More  importantly,  whereas
geographic separation forbids Eve's attack in GV from unitarily mixing
the  states $|U\rangle$ and  $|L\rangle$, in  2GV, where  the encoding
states are  based on \textit{internal} degrees of  freedom, the attack
can mix encoding states.  Thus, there is no bar on Eve's accessing the
coherence between the particles, to gain partial information about the
Bell  state  being sent  even  when  restricted  to attack  on  single
particles.   Thus,  unlike  in  GV,  there is  an  information  versus
disturbance  trade-off even  when Eve  does not  use  dummy particles,
which we discuss below.

 Lastly, our protocol satisfies the stronger QSDC security
requirement,  while  GV  in  its  original  form  is  a  protocol  for
deterministic QKD which  cannot be used for QSDC, but  can be used for
DSQC  \cite{pathak-book}.   This  can  be  understood  clearly  by
considering  that   Alice  sends  a  meaningful  message   to  Bob  by
transmitting a  sequence of $|\Psi_{0}\rangle$  and $|\Psi_{1}\rangle$
using the original  GV protocol.  In this situation,  when Alice sends
$|U\rangle$ then Eve can keep it  with her and substitute it by a fake
$|U\rangle$ and  send that to  Bob without causing any  delay.  Later,
when $|L\rangle$  is sent by Alice  then also Eve will  keep that with
her and  send a  fake $|L\rangle$ to  Bob.  Eve can  now appropriately
superpose  $|U\rangle$  and  $|L\rangle$  and  obtain  the  meaningful
information  (message)  encoded  by  Alice.  To  prevent  this,  Alice
randomizes her  transmission schedule.  Eve can  still block particles
and decode  the random  bits, but she  will be eventually  caught when
Alice and Bob  compare the sending and receiving  times.  The point is
that GV works  by \textit{streaming} qubits.  Thus by  the time she is
caught, the  encoded information will have already  been leaked.  This
leakage  is not a  problem with  GV protocol  (QKD), because  if Eve's
interference is too high, Alice and  Bob will not use that key for any
future  encryption.  In  contrast to  GV, our  protocol  uses particle
order  arrangement  in  place  of  time  schedule  randomization,  and
further, we  use \textit{block  transmission} \cite{LL02} in  place of
stream  transmission.   As a  result,  eavesdropping  does not  reveal
information  as  the  coordinates  of  the partner  particles  of  the
information encoded qubits are announced  only at the last step of the
protocol, i.e., after confirming that no eavesdropping has happened in
the   second  step  of   communication  when   $\Sigma_2^{(a|V)}$  and
$\Sigma_2^{(a|C)}$    are   communicated.    Clearly    the   proposed
cryptographic protocol is suitable DSQC.  \color{black}

Assuming  ideal sources  and detectors  with  Alice and  Bob, the  PoP
device   makes   the  protocol   exponentially   sensitive  to   Eve's
intervention.  Suppose Eve chooses to attack fraction $f$ of $n$ pairs
of  particles transmitted.  Let  $m \equiv  \lfloor  nf\rfloor$ is  an
integer.  The probability that  the $m$ particles are pair-wise closed
(i.e., every particle's twin is  within the attacked group) is $p_{\rm
  closed}  \equiv \left(  \begin{array}{c} n  \\  m \end{array}\right)
\left(  \begin{array}{c} 2n \\  2m \end{array}\right)^{-1}$  while the
probability that  all selected $m$  particles are correctly  paired by
Eve     in    the     closed    group     is    $p_{\rm     pair}    =
\frac{1}{m-1}\frac{1}{m-3}\cdots\frac{1}{3}$.   Thus  the  probability
Eve's attack produces no  error is $p_{\rm closed}p_{\rm pair}$, which
is exponentially small.

 In our protocol, the efficiency of $\frac{1}{3}$, can be improved
upon  in practice  if the  observed  noise level  remains stable  over
sufficiently many  runs, and thus  fewer quantum resources need  to be
sacrificed  to determine  it.  Our  protocol as  stated makes  no such
assumption  about  the  noise,  and  thus  considers  the  worst  case
scenario.  Consequently, in every transmission step, we have used half
of  the transmission  qubits for  error checking.   The  statistics of
random sampling then guarantees that the probability that the fraction
of errors observed in the  check bits deviates from the error fraction
in the code bits, is exponentially low \cite{NC00}. 

\section{\label{sec:Conclusions} Conclusions and Discussions}

  A  two-particle  QSDC  protocol  has  been  proposed,    with  the
motivation of understanding the  similarity and difference between the
origins of security in  GV and a multi-particle orthogonal-state based
cryptography scheme.  It may be  noted that 2GV is technically similar
to   CORE  QKD   protocol  \cite{DL03}--   with  added   ideas  (block
transmission technique) from Ref.   \cite{LL02}-- rather than to GV. A
non-trivial difference between the two situations was noted.  2GV uses
internal  degrees of  freedom, while  GV  uses the  spatial degree  of
freedom,    as   a    result   of    which   the    nature    of   the
information-vs-disturbance trade-off and  the options available to Eve
are  quite  different,  apart  from  the  obvious  difference  due  to
employing  different  numbers  of  particles.  The  PoP  technique  is
crucial to 2GV, while it can optionally be used to enhance security of
GV. However, for  GV, it suffices to increase  the parameter $\gamma$,
which is experimentally easy to implement.

\section{acknowledgment}

PY thanks the Raman Research Institute, Bangalore, India for a student
fellowship, during  which most of this  work was completed.  AP and RS
thank Department  of Science and  Technology (DST), India  for support
provided  through projects SR/S2/LOP-0012/2010  and SR/S2/LOP-02/2012,
respectively.  AP  also thanks  the Operational Program  Education for
Competitiveness - European  Social Fund project CZ.1.07/2.3.00/20.0017
of the Ministry of Education, Youth and Sports of the Czech Republic.

\bibliography{preeti}

\end{document}